\documentclass[aps,twocolumn,pre,floatfix,superscriptaddress]{revtex4-2}
\usepackage{amssymb}
\usepackage{amsmath}
\usepackage{bm}
\usepackage{graphicx}
\usepackage{dcolumn}
\usepackage{longtable}
\usepackage{color}
\usepackage[mathscr]{euscript}

\newcolumntype{C}[1]{>{\hfil}m{#1}<{\hfil}}

\begin{document}
\title{Highly fluctuating double-$q$ magnetic order in the van der Waals metal CeTe$_3$}

\author{Ryutaro~Okuma}
\email{rokuma@issp.u-tokyo.ac.jp}
\affiliation{%
 Institute for Solid State Physics, University of Tokyo, Kashiwa, Chiba 277-8581, Japan
}%
\author{Yuita Fujisawa}
\affiliation{Research Institute for Synchrotron Radiation Science, Hiroshima University, Higashi-Hiroshima 739-0046, Japan
}%
\author{Natsumi Maekawa}
\affiliation{Graduate School of Engineering and Science, University of the Ryukyus, Okinawa 903-0213, Japan}
\author{Akiko Nakao}
\affiliation{Neutron Science and Technology Center, Comprehensive Research Organization for Science and Society, Tokai, Ibaraki 319-1106, Japan}
\author{Yoshihisa Ishikawa}
\affiliation{Neutron Science and Technology Center, Comprehensive Research Organization for Science and Society, Tokai, Ibaraki 319-1106, Japan}
\author{Riki Kobayashi}
\affiliation{Faculty of Science, University of the Ryukyus, Okinawa 903-0213, Japan
}%
\author{Yoshinori Okada}
\affiliation{Quantum Materials Science Unit, Okinawa Institute of Science and Technology Graduate University, Okinawa 904-0495, Japan
}%
\author{Daichi Ueta}
\affiliation{Institute of Materials Structure Science, High Energy Accelerator Research Organization (KEK), Tsukuba, Ibaraki 305-0801, Japan
}%
\date{\today}
\begin{abstract}
CeTe$_3$ is a van der Waals antiferromagnet composed of magnetic [CeTe]$^+$ layers coupled to highly conducting Te$^{0.5-}$ square nets. Its simple quasi-two-dimensional electronic structure and cleavable nature make it an appealing platform for exploring correlated magnetism in reduced dimensions. To clarify the nature of its low-temperature state, we performed single-crystal neutron diffraction down to 0.3 K, complemented by scanning tunneling microscopy. A magnetic transition near 1.5 K gives rise to incommensurate Bragg peaks at $q_{\pm}\sim(\pm0.17,0,0.31)$, consistent with a double-$q$ magnetic order whose moments are predominantly aligned along the $c$ axis. The strongly reduced ordered moment is consistent with enhanced quantum fluctuations driven by $c$–$f$ hybridization, while the deviation of the propagation vectors from simple nesting suggests a coupling to residual charge-density-wave instabilities of the quasi-one-dimensional Te-derived bands. These results indicate that CeTe$_3$ hosts a correlated magnetic ground state where spin and itinerant charge degrees of freedom are intimately linked in the van der Waals limit.
\end{abstract}
\maketitle

\section{Introduction}
The competition between the Kondo effect and the Ruderman–Kittel–Kasuya–Yosida (RKKY) interaction in rare-earth compounds provides a central framework for understanding correlated electron phenomena\cite{doniach1977kondo}. The proximity to a quantum phase transition between a heavy Fermi-liquid state and a magnetically ordered state in the weak $c$–$f$ hybridization regime has given rise to a rich variety of emergent phases\cite{stewart1984heavy,pfleiderer2009superconducting,si2010heavy}. Most studies to date have focused on intermetallic compounds with three-dimensional connectivity\cite{kirchner2020colloquium}. In this context, the discovery of the van der Waals (vdW) heavy-fermion compound CeSiI\cite{okuma2021CeSiI,jang2022exploring,posey2024two,fumega2024nature,vijayvargia2024nematic,torres2025glassy,turkel2025nodal} has recently stimulated growing interest in extending heavy-fermion physics into the vdW regime. Owing to their intrinsic two-dimensionality and cleavable nature, vdW materials provide a versatile platform for exploring Kondo physics and quantum magnetism in the two-dimensional limit.

\begin{figure}
\includegraphics[width=8cm]{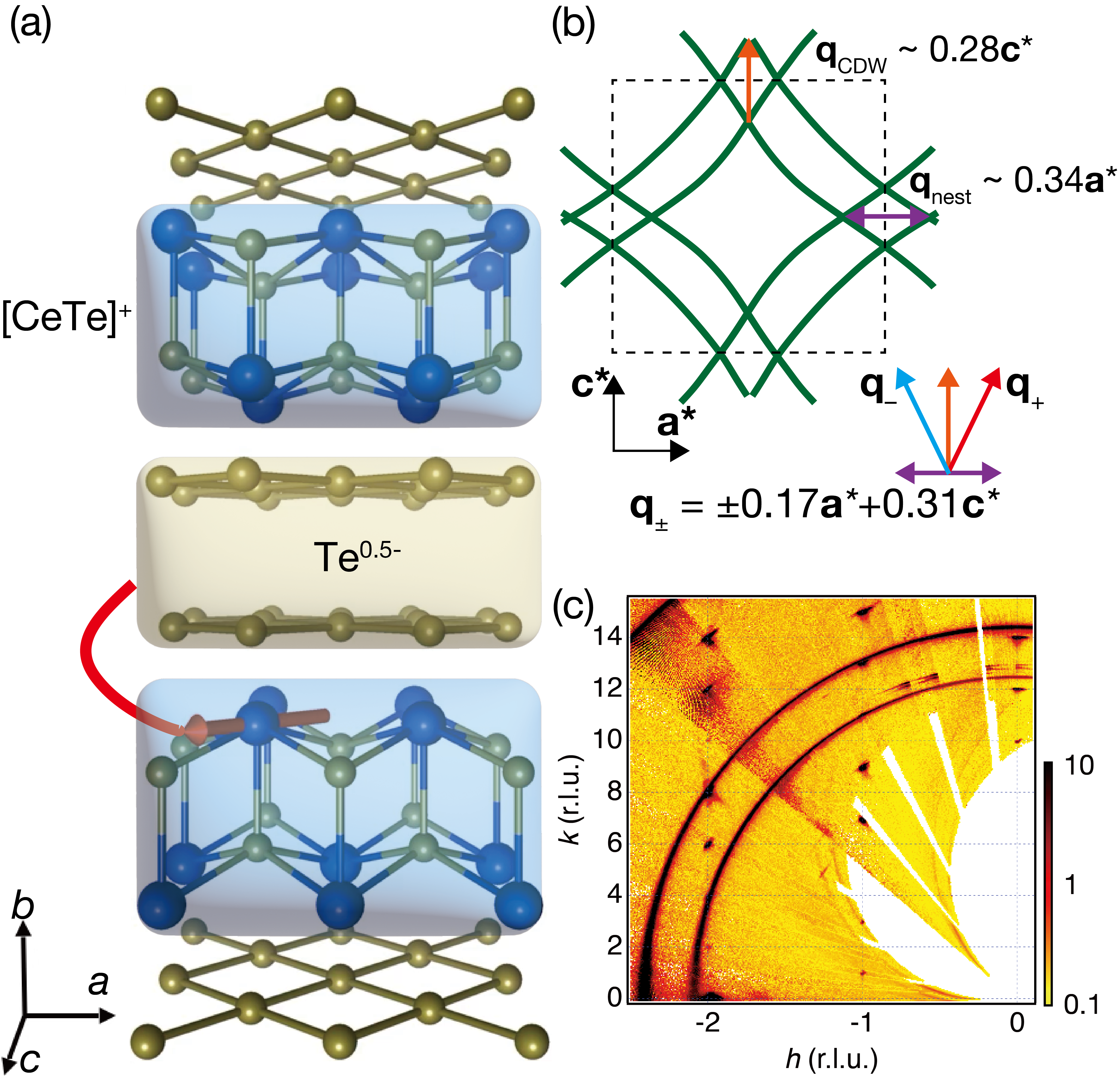}
\caption{\label{fig:overview}vdW antiferromagnet CeTe$_3$. 
(a) Crystal structure, where blue and gold spheres represent Ce and Te atoms, respectively. The conducting Te$^{0.5-}$ square nets are stacked along the $b$ axis via vdW coupling and interact with the magnetic [CeTe]$^+$ layers through Kondo coupling. Crystal structure is drawn by VESTA \cite{momma2008vesta}.
(b) Quasi-two-dimensional Fermi surface of CeTe$_3$. The nearly tetragonal slab symmetry gives rise to characteristic nesting vectors along the $a^*$ and $c^*$ axes. A CDW with $q \sim 0.28c^*$ (red vertical arrow) is already established above room temperature, while the magnetic propagation vectors identified in this work are shown by the bottom arrows having both $a^*$ and $c^*$ components. 
(c) Neutron diffraction pattern in the $hk0$ plane at 0.3 K. Sharp nuclear Bragg peaks without splitting or streaking along $K$ demonstrate the high crystalline quality of the measured single crystal. The color scale is shown to the right.}
\end{figure}

Among the few known vdW metals\cite{lin2022magnetic}, the rare-earth tritellurides $R$Te$_3$ ($R$ = lanthanide) offer a simple electronic structure and versatile magnetic properties arising from the combination of various rare-earth moments\cite{yumigeta2021advances}. The crystal structure consists of highly conducting Te$^{0.5-}$ square nets separated by magnetic [$R$Te]$^{+}$ slabs [Fig.~\ref{fig:overview}(a)]. The partially filled Te $p$ orbitals form sheet-like Fermi surfaces that undergo a charge-density-wave (CDW) transition along the $c^*$ direction above room temperature. In magnetic members such as TbTe$_3$\cite{chillal2020strongly} and DyTe$_3$,\cite{akatsuka2024non} strong coupling between the CDW and rare-earth magnetism stabilizes complex noncoplanar spin textures at low temperatures.

Within the family of $R$Te$_3$ compounds, CeTe$_3$ stands out as a particularly promising platform for studying the interplay between heavy-fermion behavior and low-dimensional quantum magnetism. The magnetic moment of Ce$^{3+}$ exhibits easy-plane anisotropy due to its nearly tetragonal crystal field\cite{ueta2022anomalous}. Previous studies have reported two successive magnetic anomalies at $T_\mathrm{N1} \approx 3$ K and $T_\mathrm{N2} \approx 1.3$ K: a broad heat-capacity peak and a weak cusp in susceptibility at $T_\mathrm{N1}$, followed by a sharp heat-capacity anomaly and a pronounced suppression of susceptibility along the $a$ and $c$ axes at $T_\mathrm{N2}$\cite{iyeiri2003magnetic,ru2006thermodynamic,okuma2020fermionic}. Application of chemical pressure realizes a magnetic order along the hard axis\cite{okuma2020fermionic,ueta2022anomalous}, suggesting unconventional magnetic fluctuations\cite{kruger2014fluctuation} in the ground state, whereas the detailed magnetic structure remains unresolved. While early transport measurements were interpreted in terms of a weak Kondo effect\cite{iyeiri2003magnetic,ru2006thermodynamic}, more recent observations of electron-mass enhancement below $T_\mathrm{N1}$ and a $-\log T$ resistivity contribution associated with Ce magnetism\cite{watanabe2021shubnikov,zeng2026kondo} suggest that CeTe$_3$ may lie near the border of a quantum-critical regime, consistent with its relatively large low-temperature heat capacity\cite{ru2006thermodynamic}. Therefore, the microscopic information on the magnetic structure is indispensable for understanding the possible heavy-fermion behavior in this quasi-two-dimensional system.

We combined single-crystal neutron diffraction and scanning tunneling microscopy (STM) to determine the magnetic structure of the van der Waals metal CeTe$_3$. The measurements reveal an incommensurate double-$q$ order with moments along the $c$ axis and a strongly reduced ordered moment accompanied by a charge modulation in STM, signaling enhanced quantum fluctuations consistent with $c$–$f$ hybridization. The deviation of the propagation vectors from simple nesting indicates coupling between magnetic and residual charge-density-wave instabilities of the quasi-two-dimensional Fermi surface.

\section{Experiment} Single crystals of CeTe$_3$ were grown using a Te flux, following the procedure described in Ref.~\cite{okuma2020fermionic}. Preliminary neutron diffraction measurements were carried out using HRC at J-PARC, PONTA at JRR-3, and WOMBAT at ANSTO.
Comprehensive single-crystal neutron diffraction experiments were subsequently performed on the time-of-flight diffractometer SENJU at J-PARC. A $^3$He cryostat enabled measurements down to 0.3 K, and data were collected at temperatures between 0.3 and 5 K. Data reduction and visualization were carried out with the STARGazer software package\cite{ohhara2016senju}.
The crystal used for neutron diffraction had dimensions of 4.25 $\times$ 0.65 $\times$ 2.91 mm$^3$ and a mass of 40.8 mg. Its quality was examined by X-ray Laue diffraction. Because the tetragonal symmetry of the CeTe$_3$ slab is broken only by the stacking sequence, stacking faults can readily induce twinning, where a minor domain is related to the major one by a 90$^\circ$ rotation about the $b$ axis. We confirmed that the crystal used for the measurement is free from such twinning, as demonstrated by the distinct reflection conditions between the $hk0$ ($h+k =$ even) and $0kl$ ($k =$ even) planes [Fig.~\ref{fig:overview}(c)].
Nuclear and magnetic structure refinements were performed using the FullProf Suite\cite{rodriguez1993recent}. Details of the nuclear refinement are provided in Appendix~\ref{A:nuclear}. The magnetic basis vectors were obtained using the SARAh webRefine program\cite{wills2025sarah}. Symmetry analysis of the magnetic structures was performed by ISODISTORT\cite{campbell2006isodisplace}.
For the scanning tunneling microscopy (STM) experiments, single crystals of CeTe$_3$ were cleaved in ultra-high vacuum at room temperature. A chemically etched W tip was calibrated on Au(111) to ensure a well-defined shape and a flat density of states. Further experimental details are described in Ref.~\cite{fujisawa2025versatile}.

\section{Results and Discussions}
\begin{figure}
\includegraphics[width=80mm]{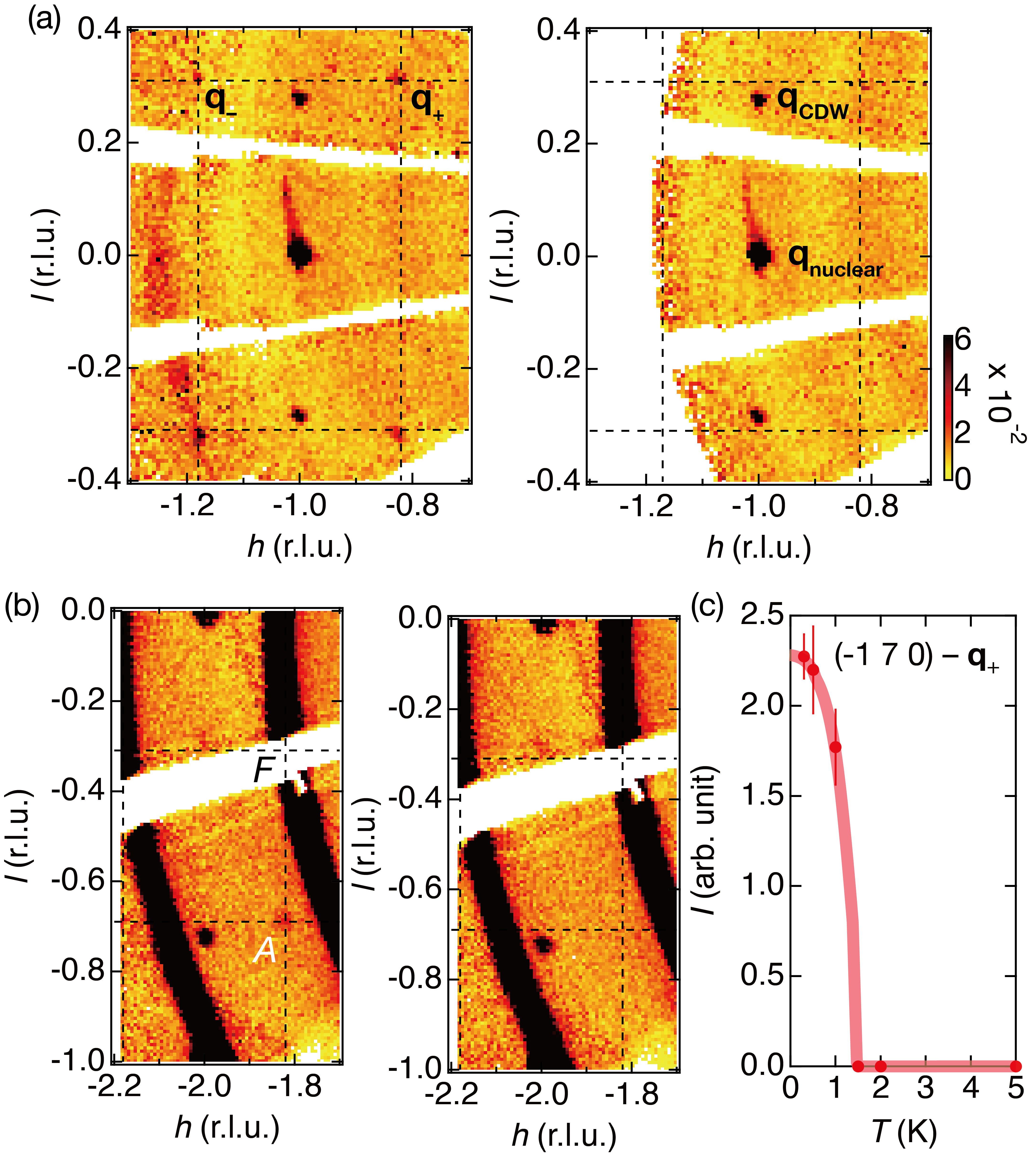}
\caption{\label{fig:propgationvevtor} 
Magnetic order in CeTe$_3$ detected by neutron diffraction. 
(a) Diffraction patterns in the $h5l$ plane at 0.3 K (left) and 5 K (right). Crossings of the dashed lines indicate the magnetic Bragg peak positions. 
(b) Diffraction patterns in the $h6l$ plane at 0.3 K (left) and 5 K (right). 
(c) Temperature dependence of the integrated intensity at $(-1,~7,~0)-q_+$. The transparent red line is a guide to the eye.}
\end{figure}

Figure~\ref{fig:propgationvevtor}(a) displays the $h5l$ plane measured at 0.3 K, where several intense magnetic Bragg peaks appear. At 5 K, well above $T_{N1}$, main nuclear Bragg peaks are observed together with satellite peaks at $q_\mathrm{CDW} = 0.28c^*$, consistent with earlier reports on the CDW order\cite{kim2006local,malliakas2006divergence}. At 2 K, well below the first transition at $T_{N1}$, the diffraction pattern still contained only structural peaks and no clear signature of diffuse scattering was observed in the current experiment. Magnetic Bragg peaks appeared only below $T_{N2} \approx 1.3$ K and all of them can be indexed by two propagation vectors, $q_+ = 0.17a^* + 0.31c^*$ and $q_- = -0.17a^* + 0.31c^*$, which are related by a mirror symmetry perpendicular to the $a$ axis in the paramagnetic state. The strongest magnetic peaks are found primarily in odd-$k$ planes and become vanishingly weak at large momentum transfers ($|Q| \gtrsim 1$~\AA$^{-1}$), as expected from the Ce$^{3+}$ magnetic form factor.

The observed diffraction pattern is consistent with either two domains of single-$q$ magnetic structures related by mirror symmetry or a double-$q$ structure formed as a linear combination of the two propagation vectors. Because no reflections arising from linear combinations of $q_+$ and $q_-$ were detected, neutron diffraction alone cannot distinguish between these two possibilities. We therefore first determined the basis vectors of each propagation vector in the double-$q$ model and then examined its validity in light of recent STM results\cite{fujisawa2025versatile}.

Details of the symmetry analysis and magnetic scattering selection rules are provided in Appendix~\ref{A:symmetry}. The irreducible representations $mM_1$ and $mM_2$ comprise two types of basis vectors, $A$ and $F$, with components along $x$, $y$, and $z$, which satisfy distinct reflection conditions. As the $A$- and $F$-type contributions are almost decoupled in the $k = 3n$ ($n$: integer) planes (see Table~\ref{tab:irreps}), the $h6l$ plane is shown in Fig.~\ref{fig:propgationvevtor}(b). A clear satellite peak at $(-2,~6,~-1)$ supports the presence of $A$-type order. Given the strong $XY$ anisotropy of Ce$^{3+}$, the observed intensity is attributed to the $x$ or $z$ component of the $A$-type order within the $mM_1$ representation.

Quantitative refinements were performed using all integrated intensities, with the ordered moments of the $q_+$ and $q_-$ components constrained to be equal. For simplicity, we first considered models with a single basis vector. Refinements assuming $F$-type magnetic orders yielded poor fits ($\chi^2 > 18$) and were therefore excluded. The best agreement was obtained for the $A_z$ order, corresponding to sinusoidally modulated moments aligned along the $c$ axis. This result is consistent with the spin-flop transition along the $c$-axis below $T_\mathrm{N2}$\cite{okuma2020fermionic}. We also tested a mixed model combining $A_z$ and $A_x$ components, which slightly improved the agreement factor. However, since the deduced canting from the $c$ axis is small, precise determination of this deviation is left for future studies.

The Fourier component of the refined ordered moment in the $A_z$ model was $0.32\mu_\mathrm{B}$, corresponding to maximum moment values of $0.64\mu_\mathrm{B}$ 
 and spatial average of the magnetic moment size $<|S|>\sim 0.32\times8/\pi^2= 0.26\mu_\mathrm{B}$ for the double-$q$ structure. The pronounced reduction from the moment size expected from the crystal field  and magnetization study, namely $0.833\mu_\mathrm{B}$\cite{okuma2020fermionic,ueta2021crystalline}, indicates strong quantum fluctuation in the ground state, much larger compared to another van der Waals Ce compound, CeSiI,\cite{okuma2021CeSiI,posey2024two} which develops nearly the full ordered moment. The observed suppression of the ordered moment is comparable to canonical itinerant Ce-based systems that become superconducting once magnetic order is suppressed. These materials host either incommensurate magnetic orders (e.g., CeRhIn$_5$\cite{bao2000incommensurate,PhysRevB.67.099903,christianson2002neutron}, CeRhSi$_3$\cite{aso2007incommensurate,ueta2021crystalline}, and CeIrSi$_3$\cite{aso2011spin,aso2012neutron,ueta2021crystalline}) or commensurate ones (e.g., CeIn$_3$\cite{benoit1980magnetic,lawrence1980magnetic} and CePd$_2$Si$_2$\cite{grier1984magnetic,van2000magnetic}).
\begin{table}
\caption{Refinement results of the $A$-type magnetic structures. The number of observed magnetic reflections is 22 for $q_+$ and 15 for $q_-$. Agreement factors ($R$ values) for $q_+$ and $q_-$ are listed as the left and right entries, respectively.}
\begin{ruledtabular}
\begin{tabular}{cccccc}
\textrm{Model} &
\textit{$M_\mathrm{refine}/\mu_\mathrm{B}$}& \textit{$R_F (\%)$}& \textit{$wR_{F^2} (\%)$}& \textit{$\chi^2$}\\
\colrule
$A_x$ & 0.39(2) & 18, 22 & 46, 42 & 6.3, 2.9 \\
$A_y$ & 0.38(3) & 33, 40 & 57, 71 & 9.2, 6.7 \\
$A_z$ & 0.32(1) & 15, 20 & 33, 36 & 3.3, 2.1 \\
$(A_x,A_z)$ & ($\mp$0.12(7),0.31(2)) & 14, 16 & 31, 29 & 3.1, 1.4\\
\end{tabular}
\end{ruledtabular}
\label{tab:Rfactors}
\end{table}

While the strongly fluctuating magnetic order in CeTe$_3$ may reflect several intertwined effects, an important question is whether the ordered state is governed mainly by local-moment interactions or by conduction electrons coupled to the Ce $4f$ moments. The latter scenario may originate from enhanced $c$-$f$ hybridization and residual Fermi-surface instabilities of the quasi-one-dimensional electronic structure, whereas the former is more naturally described by local spin interactions. These possibilities can be constrained by examining the magnetic structure in real space. In general, electronically driven order tends to favor multi-$q$ states through additional energy gain over a broader portion of the Fermi surface, whereas local spin interactions more commonly stabilize single-$q$ textures. Since the neutron diffraction data alone are consistent with both single-$q$ and double-$q$ configurations, we turn to STM measurements. Although STM is a surface-sensitive technique, the vdW crystal structure allows access to essential bulk information. Indeed, previous STM studies have revealed local quantities such as the CDW wavevector, CDW gap size, and Fermi-surface geometry in good agreement with bulk measurements \cite{ralevic2016charge,nakamura2024revealing, smith2024uncovering}. To minimize domain averaging, we focus here on a $10 \times 10$ nm$^2$ field of view, while a wider-field analysis consistent with the present results is provided in \cite{fujisawa2025versatile}.

\begin{figure*}
\includegraphics[width=17.5cm]{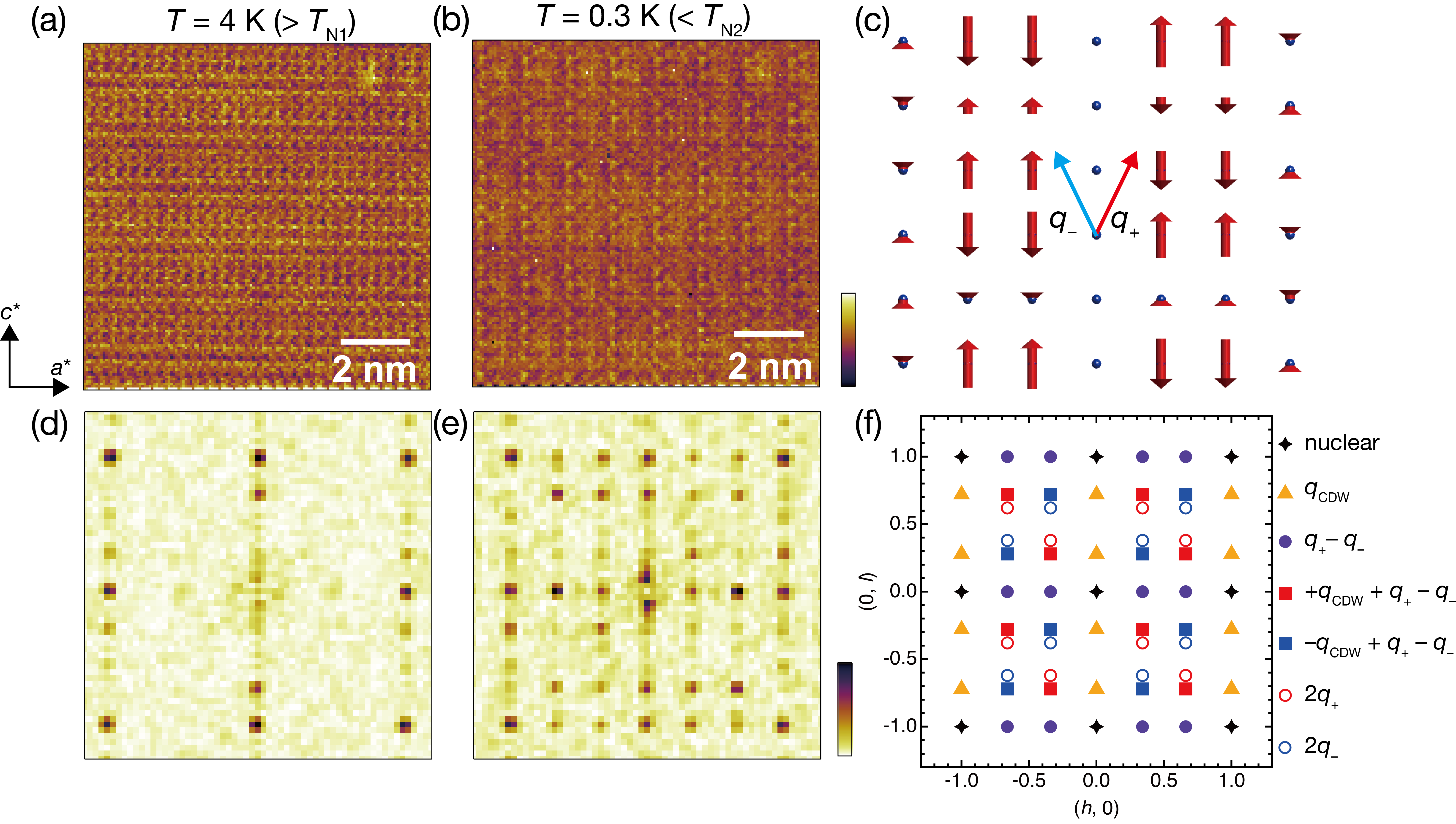}
\caption{\label{fig:stm}
(a,b) Conductance maps near the Fermi level ($E-E_\mathrm{F} = +14$ mV) measured (a) above $T_\mathrm{N1}$ and (b) below $T_\mathrm{N2}$. The set point is -60 mV/200 pA. (c) Schematic of the double-$q$ spin-density-wave (SDW) structure consistent with the neutron and STM experiments, drawn by SpinW\cite{toth2015linear}. The magnetic moment of the Ce1 site, as defined in Table~\ref{tab:BV}, is indicated. (d,e) Corresponding Fourier transform images obtained from (a) and (b), respectively. (f)Charge modulations below $T_\mathrm{N2}$ that can be explained by the double-$q$ magnetic structure and CDW order (filled symbols). Experimentally negligible peaks are shown by open symbols. See the main text for the details. Panels (a,b,d,e) are adapted from Ref.\cite{fujisawa2025versatile}.}
\end{figure*}

Figure~\ref{fig:stm}(a) shows the real-space conductance map above $T_\mathrm{N1}$, where previously reported CDW along the $c$ axis dominates the contrast in addition to the top surface Te corrugation\cite{smith2024uncovering,nakamura2024revealing}. Below $T_\mathrm{N2}$, interestingly, another charge contrast appears homogeneously in the $a$ axis as shown in Fig.~\ref{fig:stm}(b). This observation suggests a double-$q$ magnetic structure such as Fig.~\ref{fig:stm}(c), whereas a single-$q$ structure would be expected to yield strongly monoclinic electronic symmetry. In order to disentangle the periodic structures associated with the magnetic structure, Fourier transformation was performed as shown in Figs.~\ref{fig:stm}(d,e). We have also carefully assigned the observed Fourier spots as shown by the filled symbols in Fig .~\ref {fig:stm}(f), most of which are assigned by the linear combination of $q_\mathrm{CDW}$ and $q_+ – q_-$ \cite{comment_CDW3}.

Generically in the presence of non-zero spin modulation with its Fourier component given by $S_q$, and their coupling to the conduction electrons, the resulting Fourier component of charge density modulation at the momentum $q_c$ is proportional to $\delta_{q+q',q_c}S_q·S_{q'}$\cite{yasui2020imaging,hayami2021charge}.
In the case of the double-$q$ magnetic order coexisting with the high temperature CDW, charge modulations can appear generally at $m(q_+ + q_-) + n(q_+ – q_-) + lq_\mathrm{CDW} + G$, where $l$, $m$, and $n$ are integers and $G$ is an arbitrary reciprocal-lattice translational vector. 
Based on this framework, the presence of peaks associated with $q_+ - q_-$, which is forbidden in the single-$q$ structure, directly supports the formation of the double-$q$ magnetic structure.
The negligible intensity near $q_++q_-\sim0.62c^*$ and $2q_\pm\sim\pm0.34a^* + 0.62c^*$ can be attributed to the absence of nesting at these wavevectors, which is confirmed by the photoemission-derived Fermi surface and quasi-particle interference pattern in the paramagnetic state \cite{hayami2021charge, smith2024uncovering}.
It should also be noted that the slight intensity variation between $q_\mathrm{CDW}+ q_+ – q_-$ and $-q_\mathrm{CDW}+ q_+ – q_-$ is probably due to the STM tip anisotropies.
Nevertheless, the robust observation of these peaks provides evidence for the double-q magnetic structure depicted in Fig.~\ref{fig:stm}(c), and highlights a pronounced coupling between localized spins and conduction electrons, potentially involving $c-f$ hybridization.


We finally discuss the unique feature of the magnetic propagation vector in CeTe$_3$ compared to other $R$Te$_3$. The Fermi surface of the high-temperature phase without CDW is shown in Fig.~\ref{fig:overview}(b). The Fermi surface possesses two major nesting vectors, $\sim0.3c^*$ and $0.3a^*$. The well-known CDW order above room temperature develops at $q_\mathrm{CDW}\sim 0.3c^*$, while in heavier rare-earth tritellurides an additional CDW order occurs at $0.3a^*$ at lower temperatures\cite{yumigeta2021advances}.  For TbTe$_3$ and DyTe$_3$, the magnetic propagation vectors at low temperatures are $c^*/2$ and $c^*/2 - q_\mathrm{CDW}$\cite{chillal2020strongly,akatsuka2024non}. The primary component $q_\mathrm{mag}=c^*/2$ can be understood as arising from the formation of antiferromagnetic spin chains along the $c$ direction, where short-range interactions develop between large local moments in the poorly conducting $R$Te rock-salt layers. As emphasized by Aoyama\cite{aoyama2025rkky}, the high-temperature incommensurate CDW modifies the effective magnetic interactions, resulting in a pair of ordering vectors $q_\mathrm{mag}$ and $q_\mathrm{mag}-q_\mathrm{CDW}$, as realized in Tb and Dy compounds.

CeTe$_3$, however, is the only member exhibiting the magnetic propagation vector having both along the $a^*$ and $c^*$ components. The $a^*$ component of its magnetic propagation vector approximately corresponds to half of the other nesting vector, $\sim 0.34a^*$. In conventional SDW systems driven by Fermi-surface nesting, such as elemental Cr, the magnetic propagation vector coincides with the nesting vector, while the accompanying CDW forms at twice that wave vector\cite{mori1993searching}. CeTe$_3$ does not conform to this canonical SDW scenario. Instead, the residual CDW instability at $q=q_+ – q_-$ appears to drive the magnetic order, as predicted in previous studies presenting considerably good nesting conditions at the wavevector in the paramagnetic state \cite{smith2024uncovering}. This interpretation is consistent with the trend across the $R$Te$_3$ family: DyTe$_3$, TbTe$_3$, and GdTe$_3$ develop two CDWs along both the $a^*$ and $c^*$ directions, indicating that the secondary nesting instability is released through charge ordering. In contrast, CeTe$_3$ hosts only a single CDW, leaving substantial nested Fermi-surface segments, particularly along $a^*$, available to couple with Ce 4$f$ moments via $c$-$f$ hybridization, which may instead stabilize magnetic order \cite{akatsuka2024non,chillal2020strongly,raghavan2024atomic}.

We next comment on the $c^*$ component of the magnetic propagation vector ($0.31c^*$), which is similar to, but experimentally distinguishable from, that of the preexisting CDW ($q_{\mathrm{CDW}}\sim0.28c^*$) in both neutron diffraction and STM measurements \cite{fujisawa2025versatile}. While the $a^*$ component can be naturally associated with the residual nesting tendency discussed above, the additional emergence of a finite $c^*$ component is less straightforward within a simple nesting scenario. Because the primary instability along $c^*$ is already substantially relieved by the CDW formation, the magnetic ordering in this direction is also influenced by secondary effects such as exchange interactions and/or the residual susceptibility of the reconstructed Fermi surface. However, since the mismatch is small, we regard the magnetic modulation along $c^*$ as being primarily shaped by the periodic potential associated with the preexisting CDW i.e., $q_\pm\cdot c \sim q_\mathrm{CDW}\cdot c$. In this sense, the magnetic and CDW modulations along $c^*$ are distinct, yet intimately linked in origin.

While the double-$q$ magnetic order is established in this study, the direct evidence of $c–f$ hybridization is still lacking. Low energy fluctuations associated with this state could be probed by local techniques such as muon spin rotation and nuclear magnetic resonance, which may also shed light on the nature of the phase below $T_\mathrm{N1}$. Single-crystal inelastic neutron scattering would provide a direct spectroscopic probe of the magnetic excitations and could reveal characteristic avoided crossings between modes associated with $q_\pm$ peak, serving as a signature of the double-$q$ order.

\section{Summary}
In summary, we have elucidated the microscopic magnetic structure of the van der Waals rare-earth metal CeTe$_3$ using single-crystal neutron diffraction and scanning tunneling microscopy. The double-$q$ magnetic order with strongly reduced $c$-axis moments reflects pronounced quantum fluctuations, being consistent with the presence of $c$–$f$ hybridization. The deviation of the propagation vectors from simple Fermi-surface nesting indicates that the magnetic order is stabilized by a residual CDW instability of the Te-derived sheet-like Fermi surface. CeTe$_3$ thus serves as a van der Waals material where magnetic order emerges from the strong coupling between spins and a low dimensional electronic structure.

\begin{acknowledgments}
We thank K. Aoyama, Y. Niimi, T. Nakajima, and S. Hayami for helpful discussions. Neutron scattering experiments were performed on Wombat at ANSTO, Australia; SENJU and HRC at J-PARC MLF, Japan; and PONTA at JRR-3, Japan, under the General User and Neutron Scattering Program of the Institute for Solid State Physics, The University of Tokyo (Proposal No. 19508), the Institute of Materials Structure Science, KEK (Proposal Nos. 2023S01, 2024A0034), and the JRR-3 user program (Proposal No. 23401), respectively. This work was supported by JSPS KAKENHI (Grant No. 23K19027), JST ASPIRE (Grant No. JPMJAP2314), and JST PRESTO (Grant Nos. JPMJPR2591 and JPMJPR2593).
\end{acknowledgments}

\appendix
\section{Nuclear structure refinement}\label{A:nuclear}
The crystallographic information of CeTe$_3$ is summarized in Table~\ref{tab:XRD}. The intensities of the main Bragg reflections were refined using the undistorted $Cmcm$ structure without including the CDW modulation. Since the primary purpose of the refinement was to determine the scale factor for evaluating the ordered magnetic moment, we did not attempt a refinement of the incommensurately modulated structure within the superspace-group formalism, as previously reported\cite{malliakas2005square}.

\begin{table}
\caption{\label{tab:XRD}
Fractional atomic coordinates and isotropic displacement parameters of CeTe$_3$ refined from single-crystal neutron diffraction data at 5 K. Atomic coordinates and isotropic $U_\mathrm{iso}$ values (in units of 10$^{-3}$~\AA$^2$) are listed with estimated standard deviations in parentheses. The displacement parameter of all Te atoms was constrained to a common value. Space group: $Cmcm$; lattice parameters: $a = 4.3652$~\AA, $b = 25.9668$~\AA, $c = 4.3733$~\AA. Number of observed reflections: 459. Reliability factors: $R_F = 21.7\%$, $R_{F^2} = 27\%$, $wR_{F^2} = 27\%$.
}
\begin{ruledtabular}
\begin{tabular}{cccccc}
Site&
Wyckoff&
$x$&
$y$&
$z$&
$U_\text{eq}$\\
\colrule
Ce & $4c$ & 0 & 0.8315(2) & 1/4 & 1.5(12)\\
Te1& $4c$ & 0 & 0.0695(1) & 1/4 & 0.4(7)\\
Te2& $4c$ & 0 & 0.4301(1) & 1/4 & 0.4(7)\\
Te3& $4c$ & 0 & 0.7040(1) & 1/4 & 0.4(7)\\
\end{tabular}
\end{ruledtabular}
\label{tab:XRD}
\end{table}

\section{Symmetry analysis of the magnetic structure}\label{A:symmetry}
The Ce atom in CeTe$_3$ occupies single Wyckoff site $4c$ $(0,~ y,~1/4)$ in the $Cmcm$ structure, and symmetry operations produce two Ce positions in the primitive unit cell. The observed propagation vector $q_+$ is invariant by the $c$-glide operation, which also relates the Ce positions and the relative magnetic moments at these sites. The basis vectors of the magnetic moment and its reflection condition is listed in Tables \ref{tab:BV} and \ref{tab:SF}, respectively. The inversion symmetry further puts constraint on the combination on the basis vectors, which form two irreducible representations $mM_1$ and $mM_2$ as given in Table \ref{tab:irreps}. 

\begin{table}
\caption{Types of basis vectors (BVs) for the space group $Cmcm$ with $q_+ = (0.17, 0, 0.31)$ at the crystallographic $4c$ site $(0, y_\mathrm{Ce}, 1/4)$ with $y_\mathrm{Ce}$ = 0.8315. $\delta=e^{i\pi q_+\cdot c}$ is a displacement phase factor that takes into account the fact that one of the Ce positions is displaced relative to the other by $c/2$ in the direction of the propagation vector $q_+$. As $q_+$ and $q_-$ have the same $c^*$ component, BVs are the same. }  
\begin{ruledtabular}
\begin{tabular}{cccccc}
\textrm{Ce site}&
\textit{x}&
\textit{y}&
\textit{z}&
\textit{F}&
\textit{A}\\
\colrule
Ce1&0&$y_\mathrm{Ce}$&1/4&1&1\\
Ce2&0&$1-y_\mathrm{Ce}$&3/4&$\delta$&$-\delta$
\end{tabular}
\end{ruledtabular}
\label{tab:BV}
\end{table}

\begin{table}
\caption{Structure factor for the basis vectors defined in Table \ref{tab:BV}. The magnetic structure factor is defined as
$F(Q) = F((h,k,l)\pm q_i) = f_C \sum_{i=\pm,n=1,2} M_{q_i,n}\, e^{iQ \cdot r_n}$, where $Q$ is the momentum transfer,$q_i$ denotes the magnetic propagation vector, $M_{q_i,n}$ is the Fourier component of the magnetic moment at the $n$-th Ce site, and $r_n$ is its position. The factor $f_C = 1+e^{i\pi(h+k)}$ accounts for the $C$-centering of the unit cell. Because the Ce position is close to $y_\text{Ce}\sim 5/6$, the structure factor for the $F$-type component vanishes when $k$ is a multiple of 3 and $l$ is odd, whereas the $A$-type contribution vanishes under the same condition when $l$ is even.}
\begin{ruledtabular}
\begin{tabular}{cc}
\textrm{BV}&
$F((h,k,l)+q_i)$\\
\colrule
$F$&$2f_C\cos(2\pi(ky_\text{Ce} + l/4))$\\
$A$&$2f_C\sin(2\pi(ky_\text{Ce} + l/4))$
\end{tabular}
\end{ruledtabular}
\label{tab:SF}
\end{table}

\begin{table}
\caption{Irreducible representations (IRs) constructed from the basis vectors given in Table \ref{tab:BV}.$A_\mu$ and $F_\mu$ $(\mu=x,y,z)$ indicate the $\mu$ component of each basis vector.}  
\begin{ruledtabular}
\begin{tabular}{cc}
\textrm{Irreps}&
\textit{BV}\\
\colrule
$mM_1$&$A_x, iF_y, A_z$\\
$mM_2$&$iF_x, A_y, iF_z$
\end{tabular}
\end{ruledtabular}
\label{tab:irreps}
\end{table}

\bibliography{ref_arxiv}
\end{document}